\documentclass[11pt]{article}
\usepackage{epsfig,amssymb,graphicx,amsmath,amsthm,subcaption}
\usepackage{tikz}
\usepackage{float}
\usetikzlibrary{shapes.geometric, arrows}
\usetikzlibrary{positioning}
\usepackage{enumerate}
\usepackage{calrsfs}

\def\tanh{\mathop{\textrm{tanh}}\nolimits}
\def\arctanh{\mathop{\textrm{arctanh}}\nolimits}

\begin{document}
\title{Mathematical model for the impact of awareness on the dynamics of infectious diseases}
 	
	\author{G.O. Agaba,\hspace{0.5cm}Y.N. Kyrychko,\hspace{0.5cm}K.B. Blyuss\thanks{Corresponding author. Email: k.blyuss@sussex.ac.uk} 
\\\\ Department of Mathematics, University of Sussex, Falmer,\\
Brighton, BN1 9QH, United Kingdom}

\maketitle	

\begin{abstract}

This paper analyses an SIRS-type model for infectious diseases with account for behavioural changes associated with the simultaneous spread of awareness in the population. Two types of awareness are included into the model: private awareness associated with direct contacts between unaware and aware populations, and public information campaign. Stability analysis of different steady states in the model provides information about potential spread of disease in a population, and well as about how the disease dynamics is affected by the two types of awareness. Numerical simulations are performed to illustrate the behaviour of the system in different dynamical regimes.
\end{abstract}

\section{Introduction}

The last two decades have witnessed a number of major outbreaks of infectious diseases, including swine flu, SARS, Ebola, and, most recently, the Zika virus. Due to the globalised travel and significant advances in social media, information about these outbreaks is spreading quite quickly, and this, in turn, can have a profound effect on the actual epidemic dynamics \cite{Ferg07,Jones09,Nishiura,Pruyt}. Interestingly, awareness can have very complex and sometimes unexpected effects on the dynamics of the disease spread. It can have a clearly positive influence, where disease propagation is minimised or fully stopped by various disease control measures, such as the use of face masks, condoms or other tools appropriate for specific diseases, as well as vaccination and even quarantine, with examples ranging from the plague outbreak in the English village of Eyam in 1665-1666 \cite{Mur}, where the village completely sealed itself off to prevent further transmission of plague, to more recent outbreaks of swine influenza \cite{Jones09} and Ebola \cite{Pruyt}. On the other hand, the spread of information about a disease can also result in anxiety and panic, which can lead to undesired consequences, such as the uncontrolled spread of plague during the 1994 outbreak in one of the states in India, where by fleeing the endemic area the people carried the disease with them, thus infecting other parts of the country \cite{Ramali}. Another example is an HPV vaccination campaign in Romania, which has failed largely due to a very low take-up of vaccination mostly resulting from the negative press coverage \cite{penta}. In light of this complexity of behavioural changes in the population in the presence of awareness, it is important to understand how the concurrent spread of disease and awareness affects disease dynamics.

A number of mathematical models have been proposed to analyse the effects of information and awareness on the spread of epidemics. These models can be roughly divided into two major classes in terms of how they mathematically represent populations of interacting individuals: network-based models, where individuals are represented as network nodes, and edges correspond to possible connections along which a disease can be transmitted \cite{Funk09,Funk10,gross,Hatzopoulos,Juher,Sahneh11,Wang13,Wu12}, and mean-field models that assume global mixing between individuals in the population (see Manfredi and d'Onofrio \cite{Man} for a review of some of the existing models). Funk et al. \cite{Funk09} have investigated how the spread of awareness prompted by a first-hand contact with the disease affects the spread of the disease. They showed that in a social network, the spread of awareness and the resulting reduction in susceptibility does not only lower the incidence of the disease, but in some cases can even prevent onset of epidemics, thus implying that awareness can act as an effective measure of disease control. Furthermore, their results suggest that in the presence of an infectious disease, social distancing should be considered not only from the perspective of some centrally controlled action, but also in terms of self-initiated behavioural changes of individuals. This is further supported by Kleczkowski et al. \cite{Kleczkowski} who analysed two dimensions of behavioural changes: reduction in the number of contacts (staying at home) and reduction in the likelihood of contacts resulting in infections (washing hands). Their results revealed that ``washing hands" strategy appeared to be more effective for short-lived diseases while ``staying at home" was better for long-lived diseases.

Mean-field models have provided an alternative approach for modelling the effect of awareness on disease transmission. One possibility is to represent awareness as the reduction of transmission (contact) rate by some factor that grows with the number of infected individuals, with the common choices being either a saturated \cite{Cui1,LiCui,Sun,Tch1} or exponential \cite{Cui2,Liu2,Tch2} growth of the reduction factor. In the specific context of STIs, most individuals are actually aware of the spreading infection but they may still choose not to respond to the threat, so Kiss et al. \cite{Kiss10} have considered the effects of disease awareness in the case of STIs, where the rate of information transmission has the form of a saturating function of the number of infected individuals, and the value of information is allowed to decay over time. The authors have shown that whilst the population-wide awareness does not affect the epidemic threshold, it acts to reduce the infection prevalence at endemic equilibrium. Another approach is to introduce a separate compartment for the ``media" variable that effectively represents the level of awareness in the population, and the populations move from the unaware to aware compartments at rates proportional to this level of awareness \cite{Misra11,Misra11b,Misra15,Samanta13}. Mean-field models have highlighted a number of important features of dynamics associated with the simultaneous spread of disease and awareness, such as the occurrence of multiple disease outbreaks due to the spread of information \cite{Liu2}, co-existence of multiple feasible equilibria \cite{Cui2,Liu2}, behavioural changes that are dependent on disease prevalence \cite{Bauch13,don09,Perra11,Poletti09}. They have also helped analyse optimal disease control programs \cite{LiCui,Roy15,Tch1,Wang16} and the role of time delay in the response to awareness campaigns on disease dynamics \cite{Greenhalgh15,Zhao14,Zuo14,Zuo15}.

In this paper, we focus on the question of how the dissemination of {\it private awareness} arising from direct contacts between unaware and aware individuals, and {\it public awareness} stemming from population-wide information campaigns affect the dynamics of the disease spread \cite{Bass69}. The model includes the possibility of direct contacts between unaware and aware individuals regardless of their disease status, and it also takes into account public spread of awareness through various media and information campaigns.

The outline of the paper is as follows. In the next Section we derive the model and discuss its basic properties. Section 3 contains the analysis of feasibility and conditions for stability of different steady states. In Section 4 we explore how different aspects of disease awareness affect epidemic threshold, and also present numerical simulations of the model to illustrate different dynamical behaviours. The paper concludes in Section 5 with the discussion of results and future outlook.

\section{Model derivation}

In order to analyse the effects of awareness on the dynamics of a directly transmitted disease, we use an SIRS-type model similar to \cite{Funk10b}, and divide the overall population into two major compartments: unaware susceptible, infected and recovered individuals (denoted by $S_n$, $I_n$ and $R_n$) and aware susceptible, infected and recovered individuals (denoted by $S_a$, $I_a$ and $R_a$). The need to include infected and recovered individuals who are themselves unaware of the epidemic stems from the observation that many infectious diseases possess non-negligible incubation periods, during which they are asymptomatic, or symptoms may actually never develop at all, and hence, individuals may be completely unaware that they are actually the carriers of infections. Notable examples of such infections include tuberculosis  and many STIs, including chlamydia, gonorrhoea and HIV \cite{keeling,nelson}.

A disease is characterised by a transmission rate $\beta$ for unaware population, which is reduced by the factors $ 0 < \sigma_i\leq 1$ and $0 < \sigma_s\leq 1$ that represent the decrease in infectivity and susceptibility, respectively. A reduction in infectivity occurs due to infected individuals taking treatment or possibly staying at home (quarantine) to reduce their contacts, while a reduction in susceptibility is associated with susceptible individuals taking measures for disease prevention, such as face masks, vaccination or tablets etc. Infected individuals recover at a rate $r$, which is further amplified by a factor $\varepsilon$ for aware individuals. Upon recovery, it is assumed that individuals remain immune to the disease for an average period of $1/\delta$, after which time they return to their respective class of susceptibles. The duration of this temporary immunity for aware individuals is taken to be longer by a factor of $1/\phi$ \cite{Funk10b}.

As mentioned in the Introduction, there are several different ways how disease awareness can be incorporated into the mean-field model. Irrespective of whether awareness is modelled explicitly as a separate compartment, or acts as a direct modification of the disease transmission rate, a number of authors have explicitly included prevalence-dependent reduction in the disease transmission rate to signify the fact that a higher overall number of infected individuals results in a higher level of awareness \cite{Bauch13,don09,Perra11,Poletti09}. We adopt a slightly different approach used by Funk et al. \cite{Funk10b}, where we rather explicitly introduce distinct compartments for unaware and aware individuals in each of the disease states, and transitions between respective unaware and aware compartments take place at constant rates. `Private' awareness is assumed to spread from the aware section of the population to the unaware at a rate $\alpha_j$ and to be lost at a rate $\lambda_j$, where $j = 1, 2, 3$ corresponds to the susceptible, infected and recovered individuals, respectively. When compared to the model studied in Perra et al. \cite{Perra11}, this is fully analogous to a prevalence-dependent transmission of awareness. Besides this private awareness associated with direct contacts between unaware and aware individuals, we also include a possibility of a `public' or population-wide campaign aimed at reducing the impact of the disease by distributing information about this disease. Formally, this is represented in the model by direct transitions from each unaware population to an associated aware population, i.e. from $S_n$ to $S_a$, from $I_n$ to $I_a,$ and from $R_n$ to $R_a$, at a rate $\omega_j$, $j = 1, 2, 3$. With the above assumptions, the model for the simultaneous spread of the disease and awareness takes the form
\begin{equation}\label{eqn1}
\begin{array}{l}
\displaystyle{\frac{dS_n}{dt}=-\frac{(I_n+\sigma_i\,I_a)\,\beta\,S_n}{N}-\frac{\alpha_1\,(S_a+I_a+R_a)\,S_n}{N}+\lambda_1\,S_a+\delta\,R_n-\omega_1\,S_n,}\\\\
\displaystyle{\frac{dI_n}{dt}=\frac{(I_n+\sigma_i\,I_a)\,\beta\,S_n}{N}-\frac{\alpha_2\,(S_a+I_a+R_a)\,I_n}{N}+\lambda_2\,I_a-r\,I_n-\omega_2\,I_n,}\\\\
\displaystyle{\frac{dR_n}{dt}=-\frac{\alpha_3\,(S_a+I_a+R_a)\,R_n}{N}+\lambda_3\,R_a-\delta\,R_n+r\,I_n-\omega_3\,R_n,}\\\\
\displaystyle{\frac{dS_a}{dt}=-\frac{(I_n+\sigma_i\,I_a)\,\sigma_s\,\beta\,S_a}{N}+\frac{\alpha_1\,(S_a+I_a+R_a)\,S_n}{N}-\lambda_1\,S_a+\phi\,\delta\,R_a+\omega_1\,S_n,}\\\\
\displaystyle{\frac{dI_a}{dt}=\frac{(I_n+\sigma_i\,I_a)\,\sigma_s\,\beta\,S_a}{N}+\frac{\alpha_2\,(S_a+I_a+R_a)\,I_n}{N}-\lambda_2\,I_a-\varepsilon\,r\,I_a+\omega_2\,I_n,}\\\\
\displaystyle{\frac{dR_a}{dt}=\frac{\alpha_3\,(S_a+I_a+R_a)\,R_n}{N}-\lambda_3\,R_a-\phi\,\delta\,R_a+\varepsilon\,r\,I_a+\omega_3\,R_n.}
\end{array}
\end{equation}
We assume that in this model the relations $\alpha_2 \geq \alpha_3 \geq \alpha_1$ and $\omega_2 \geq \omega_3 \geq \omega_1$ hold to represent the fact that through their exposure and development of symptoms infected individuals are more likely to look for information about the disease either through their contacts or more generally in the media, and the same applies to recovered individuals, though to a smaller degree, while susceptibles are least likely to be interested in the potential outbreak. On the other hand, having become aware, we assume that susceptibles are most likely to lose their awareness as something unimportant and not directly relevant, whereas recovered and infected individuals will retain the awareness for a greater amount of time, as represented by $\lambda_1 \geq \lambda_3 \geq \lambda_2$.

\begin{figure}
\begin{center}
\vspace{-0.4cm}

\tikzstyle{unaware}=[circle, draw=black!50, fill=blue!20, thick, inner sep = 1mm]
\tikzstyle{aware}=[circle, draw=black!50, fill=red!20, thick, inner sep = 1mm]
\tikzstyle{one}=[->>, dashed, thick, left, text centered]
\tikzstyle{two}=[->>, thick, above, auto, text centered]
\tikzstyle{three}=[->, thick, above, text centered]

\begin{tikzpicture}
\node[unaware] (Sn) at (1,4) {$S_n$};
\node[unaware] (In) at (4,4) {$I_n$};
\node[unaware] (Rn) at (7,4) {$R_n$};
\node[aware] (Sa) at (1,1) {$S_a$};
\node[aware] (Ia) at (4,1) {$I_a$};
\node[aware] (Ra) at (7,1) {$R_a$};

\path 	(Sn)	edge	[two] node {$\scriptstyle \beta,  \hspace{0.2cm}  \sigma_i \beta$} (In)
		edge [one, bend right=45] node {$\alpha_1$} (Sa)
		edge [three, left] node {$\omega_1$} (Sa)
	(In)	edge [three] node {$r$} (Rn)
		edge [three, left] node {$\omega_2$} (Ia)
		edge [one, bend right=45] node {$\alpha_2$} (Ia)
	(Rn)	edge [three, below, bend right=45] node {$\delta$} (Sn)
		edge [one, bend right=45] node {$\alpha_3$} (Ra)
		edge [three, left] node {$\omega_3$} (Ra)
	(Sa)	edge [two] node {$ \scriptstyle \sigma_s \beta, \hspace{0.2cm}  \sigma_i \sigma_s \beta$} (Ia)
		edge [three, left, bend right=45] node {$\lambda_1$} (Sn)
	(Ia)	edge [three] node {$\varepsilon r$} (Ra)
		edge [three, left,  bend right=45] node {$\lambda_2$} (In)
	(Ra)	edge [three, left, bend right=45] node {$\lambda_3$} (Rn)
		edge [three, bend left=45] node {$\phi \delta$} (Sa);

\end{tikzpicture}
\caption{Model diagram: dynamics of transitions in model (\ref{eqn1}). Solid lines represent transitions associated with individuals. Arrows represent a type of ``possible transitions": double-head arrows indicate processes subject to contacts associated with the disease (solid lines) or awareness (dash lines), single-head arrows indicate processes that are not subject to contact.}\label{fig1}
\end{center}
\end{figure}
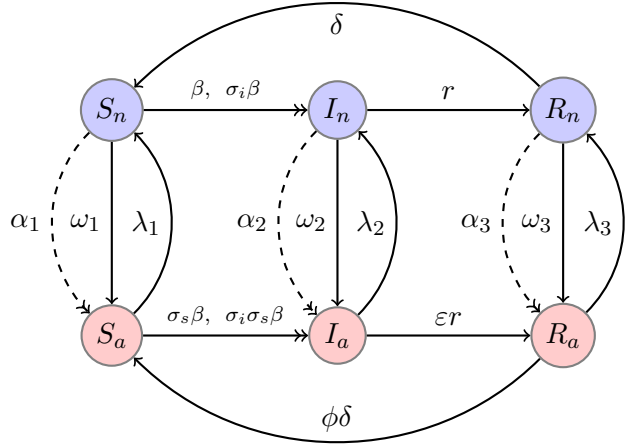

This model generalises an earlier work of Funk et al. \cite{Funk10b} by allowing the unaware susceptible and recovered populations to acquire information through public awareness programme without the need for contacts with aware individuals. This provides a very important practical difference, since preventing the disease through an appropriate information programme is very effective and more economical than treating the disease once it takes off in the population. Figure~\ref{fig1} shows the model diagram with all the transitions between different model compartments.

Since the model (\ref{eqn1}) does not include vital dynamics and there are no disease-induced deaths, this implies that the total population $N(t)=N_n(t)+N_a(t)=N$ is constant, where $N_n(t)=S_n(t)+I_n(t)+R_n(t)$ and $N_a(t)=S_a(t)+I_a(t)+R_a(t)$ are total populations of unaware and aware individuals, respectively. It is easy to show that the model (\ref{eqn1}) is well-posed, i.e. its solutions are non-negative for all $t\geq 0$.

Summing up the last three equations in (\ref{eqn1}) under the assumption of equal rates of awareness gain and loss, i.e. for $\alpha_j=\alpha,$ $\omega_j=\omega$ and $\lambda_j=\lambda$, $j = 1,2,3,$ and the fact that $N_a(t)=N-N_n(t)$ yields a logistic-type equation for the total aware population
\begin{equation}
\label{eqn4}
\frac{dN_a}{dt} = b\,N_a\,\left(1 -\frac{N_a}{K}\right)+\omega\,N_a,
\end{equation}
where $b = \alpha - \lambda - \omega$ and $ K = \frac{b\,N}{\alpha}$. This equation can be solved with the initial condition $N_a(0)=0$ to give
\[
\displaystyle{N_a(t)=\frac{N}{2\alpha}\left[b+\sqrt{b(4p+b)}\tanh\left(\frac{t}{2}\sqrt{b(4p+b)}-\arctanh\frac{b}{\sqrt{b(4p+b)}}\right)\right],\hspace{0.2cm}p=\frac{\alpha\,\omega}{b}.}
\]
From this expression it immediately follows that
\[
N_{a}(\infty)=N\, \left[\frac{1}{2}\, \left(1-\frac{\lambda+\omega}{\alpha}\right)+ \sqrt {\frac{1}{4}\,\left(1-\frac{\lambda+\omega}{\alpha}\right)^2 + \frac{\omega}{\alpha}} \right],
\]
and
\[
N_n(\infty)=N-N_{a}(\infty)=N\, \left[\frac{1}{2}\, \left(1+\frac{\lambda+\omega}{\alpha}\right)-\sqrt {\frac{1}{4}\,\left(1-\frac{\lambda+\omega}{\alpha}\right)^2 + \frac{\omega}{\alpha}} \right],
\]
This implies that as $t \to \infty$, $N_n(t)$ and $N_{a}(t)$ tend to some constant values that only depend on the rates of gain of private and public awareness ($\alpha$ and $\omega$) and the loss rate of awareness $\lambda$, but are independent of the initial conditions for individual populations or the characteristics of the disease, such as the transmission rate, or the durations of recovery or temporary immunity.

\section{Steady states and their stability}

As a first step in the analysis, we look at possible steady states of the model (\ref{eqn1}). In the absence of public awareness, i.e. for $\omega_1=\omega_3=\omega_3=0$, the model (\ref{eqn1}) always has a disease-free steady state
\begin{equation}
E_0=(S_n^*, 0, 0, 0, 0, 0)=(N,0,0,0,0,0).
\end{equation}
It can also have an {\it awareness-endemic} equilibrium (characterised by only private dissemination of awareness)
\begin{equation}
E_{0}^{0}=(S_n^*, 0, 0,S_a^*, 0, 0) \qquad \text{with} \quad S_n^* = N\,\frac{\lambda_1}{\alpha_1}, \quad S_a^* = N\,\left(1 - \frac{\lambda_1}{\alpha_1} \right),
\end{equation}
and a {\it disease-endemic} steady state
\begin{equation}
E_{1}=(S_n^*, I_n^*, R_n^*, 0, 0, 0),
\end{equation}
where
\[
S_n^* = \frac{ r\,N}{\beta}, \qquad I_n^* = \frac{\delta\,N}{\delta + r}\,\left(1 - \frac{r}{\beta}\right) \qquad \text{and} \qquad R_n^* = \frac{r\,N}{\delta + r}\,\left(1 - \frac{r}{\beta}\right).
\]
While the disease-free steady state exists for any values of system parameters, the endemic states $E_{0}^{0}$ and $E_{1}$ are only biologically feasible, provided the conditions
\begin{equation}
R_0^a=\frac{\alpha_1}{\lambda_1} > 1,
\end{equation}
and
\begin{equation}
R_0^d=\frac{\beta}{r}>1,
\end{equation}
respectively, hold.

In the case where public awareness is present, i.e. $\omega_j>0$, the disease-free steady state is actually an awareness-endemic equilibrium
\begin{equation}
E_0^{\omega}=(S_n^*, 0, 0, S_a^*, 0, 0),\hspace{0.5cm}\mbox{with}\hspace{0.5cm}S_a^* = Nh, \qquad S_n^* = N(1 - h),
\end{equation}
where
\begin{equation}\label{h_def}
\displaystyle{h = \frac{1}{2}\,\left(1-\frac{\lambda_1+\omega_1}{\alpha_1}\right)+\sqrt{\frac{1}{4}\,\left(1-\frac{\lambda_1+\omega_1}{\alpha_1}\right)^2+\frac{\omega_1}{\alpha_1}},\hspace{0.5cm}0<h<1.}
\end{equation}

The fully endemic steady state for arbitrary positive values of $\alpha_j$, $\omega_j$ and $\lambda_j$ cannot be found in a closed form, but in the particular case of equal rates of awareness spread and loss, i.e. for $\omega_j = \omega$, $\alpha_j= \alpha$ and $\lambda_j= \lambda$, it can be readily found as follows
\[
E_2^{\omega}=(S_n^*, I_n^*, R_n^*, S_a^*, I_a^*, R_a^*),
\]
with individual components being explicitly given by
\begin{equation}
\begin{array}{l}
\displaystyle{S_n^* = \frac{B \pm \sqrt{B^2 - 4\,A\,C}}{2\,A}, \qquad I_n^* = \frac{m_2\,m_7\,[N\,(1 - h) - S_n^*]}{m_2\,m_7 + r\,(\lambda + \phi\,\delta)\,m_7 + \lambda\, \varepsilon\,r\,m_6},}\\\\
\displaystyle{R_n^* =   \frac{[r\,(\lambda + \phi\,\delta)\,m_7 + \lambda\, \varepsilon\,r\,m_6]\,[N\,(1 - h) - S_n^*]}{m_2\,m_7 + r\,(\lambda + \phi\,\delta)\,m_7 + \lambda\, \varepsilon\,r\,m_6},}\\\\
\displaystyle{S_a^* = \frac{N\,m_3 - m_4\,S_n^*}{m_5}, \qquad I_a^* = \frac{m_2\,m_6\,[N\,(m_5\,h - m_3) + m_4\,S_n^*]}{m_5\,[m_6\,(m_1 + m_2) + r\,(\alpha\,h + \omega)\,m_7]},}\\\\
\displaystyle{R_a^* =  \frac{[r\,(\alpha\,h+\omega)\,m_7 + m_1\,m_6]\,[N\,(m_5\,h - m_3) + m_4\,S_n^*]}{m_5\,[m_6\,(m_1 + m_2) + r\,(\alpha\,h + \omega)\,m_7]},}
\end{array}
\end{equation}
where
\[
\begin{array}{l}
m_1= \varepsilon\,r\,(\alpha\,h+\delta+\omega),\quad m_2 = \lambda\,\delta + \phi\,\delta\,(\alpha\,h+\delta+\omega), \quad
m_3 = \lambda\,r + \varepsilon\,r\,(\alpha\,h + r + \omega),\\\\
m_4 = \beta [\sigma_i (\alpha h + \omega) + \lambda + \varepsilon r],\quad m_5=\beta\sigma_s [\sigma_i (\alpha\,h + r + \omega) + \lambda],\\\\
m_6=N(\alpha\,h + r + \omega) - \beta S_n,\quad m_7=N\lambda + \sigma_i \beta S_n,
\end{array}
\]
and
\[
\begin{array}{l}
\displaystyle{A = \beta\,[m_4\,\lambda\,\varepsilon\,r + m_5\,(m_1 + m_2)] - \beta\,\sigma_i\,[m_4\,(m_2 + \lambda\,r + r\,\phi\,\delta) + r\,(\alpha\,h + \omega)\,m_5],}\\\\
\displaystyle{B = N\,\Big(m_4\,\lambda\,(m_2 + m_3 + r\,\phi\,\delta) +  m_5\,(m_1 + m_2)\,[\beta\,(1 - h) + \alpha\,h + r + \omega]}\\
\displaystyle{\hspace{1cm}+r\,(\alpha\,h + \omega)\,m_5\,[\lambda - \sigma_i\,\beta\,(1 - h)] - \beta\,(m_3 - m_5\,h)\,[\sigma_i\,(m_2 + \lambda\,r + r\,\phi\,\delta) - \lambda\,\varepsilon\,r]\Big),}\\\\
\displaystyle{C = N^2\,\Big[ m_5\,(1 - h)\,[(\alpha\,h + r + \omega)\,(m_1 + m_2) + \lambda\,r\,(\alpha\,h + \omega)]}\\
\displaystyle{\hspace{1cm}+ \lambda\,(m_3 - m_5\,h)\,(m_2 + m_3 + r\,\phi\,\delta)\Big].}
\end{array}
\]
The endemic steady state $E_2^{\omega}$ is only feasible when the value of $S_n^*$ lies within the interval
\[
\frac{N\,(m_3 - m_5\,h)}{m_4} < S_n^* < \min \left\{\frac{N\,m_3}{m_4},\frac{N\,(\alpha\,h + r + \omega)}{\beta},N\,(1 - h)\right\},
\]
which ensures that all steady-state variables have positive values.

To analyse the stability of different steady states, we start by considering the case $\omega_1=\omega_3=\omega_3=0$ and linearise the system (\ref{eqn1}) near the disease-free steady state $E_0$. This gives a characteristic equation for eigenvalues $\mu$, which can be factorised as follows
\[
\mu(\mu + \lambda_1 - \alpha_1)(\mu + r -\beta)(\mu + \lambda_2 +\varepsilon\,r)(\mu + \delta)(\mu + \lambda_3 + \phi\delta) = 0,
\]
suggesting that the steady state $E_0$ is linearly asymptotically stable, provided
\[
\beta<r,\hspace{0.5cm}\mbox{and}\hspace{0.5cm}\alpha_1<\lambda_1,
\]
or, equivalently,
\begin{equation}\label{E0stab}
R_0^d<1,\hspace{0.5cm}R_0^a<1.
\end{equation}
Similarly, one can show that the awareness-endemic equilibrium $E_{0}^{0}=(S_n^*, 0, 0,S_a^*, 0, 0)$ is linearly asymptotically stable if
\begin{equation}\label{R0d_def}
R_0^d < \psi_0,\hspace{0.5cm}R_0^a>1,
\end{equation}
where
\begin{equation}
\psi_0= \frac{\alpha_1 [\alpha_1 \lambda_2 + \varepsilon (\alpha_2(\alpha_1 - \lambda_1) + \alpha_1 r)]}{(\alpha_1 - \lambda_1)\left[\sigma_i \sigma_s (\alpha_2(\alpha_1 - \lambda_1) + \alpha_1 r) + \alpha_2 \lambda_1 \sigma_i + \alpha_1 \lambda_2 \sigma_s \right] + \alpha_1 \lambda_1 (\lambda_2 +\varepsilon r)},
\end{equation}
whereas the disease-endemic steady state $E_1=(S_n^*, I_n^*, R_n^*, 0, 0, 0)$ is linearly asymptotically stable whenever the following conditions hold
\begin{equation}
R_0^d>1,\hspace{0.5cm}R_0^a<1.
\end{equation} 

In the presence of public awareness, i.e. for $\omega_j > 0$, linearisation near the awareness-endemic equilibrium $E_0^{\omega}=(S_n^*, 0, 0, S_a^*, 0, 0$) yields the following characteristic equation
\[
\mu\,(a_3- a_2 + \lambda_1 + \mu)(\mu^2 + \mu\,g_1 + g_2)(\mu^2 + \mu\,g_3 + g_4)= 0,
\]
where
\[
\begin{array}{l}
\displaystyle{a_1=\frac{\beta S_n^*}{N}, \hspace{0.4cm} a_2=\frac{\alpha_1 S_n^*}{N}, \hspace{0.4cm} a_{3}=\frac{\alpha_1 S_a^*}{N}+\omega_1,\hspace{0.4cm}a_4=\frac{\sigma_i\beta S_n^*}{N},\hspace{0.4cm}a_5=\frac{\sigma_s\beta S_a^*}{N}}, \\\\
\displaystyle{a_6=\frac{\sigma_i\sigma_s\beta S_a^*}{N}},
 \hspace{0.4cm} a_{7}=\frac{\alpha_2 S_a^*}{N}+\omega_2, \hspace{0.4cm} a_{8}=\frac{\alpha_3 S_a^*}{N}+\omega_3,\\\\
g_1=\lambda_3 + \phi\,\delta + a_{8} +\delta, \hspace{0.3cm} g_2= \lambda_3\,\delta + \phi\,\delta\,(a_{8} + \delta),\hspace{0.3cm} g_3=\lambda_2 +\varepsilon\,r - a_6 + a_{7} + r - a_1, \\\\
g_4=(\lambda_2 +\varepsilon\,r - a_6)\,(a_{7} + r - a_1) - (a_{7}+ a_5)\,(a_4+\lambda_2).
\end{array}
\]
It is straightforward to show that all roots of this equation (except for $\mu=0$) have negative real part, provided
\begin{equation}
\frac{\lambda_1+\omega_1}{\alpha_1}>1-2h,
\end{equation}
and
\[
R_0^d<\psi, \hspace{0.2cm} \mbox{where} \hspace{0.2cm}\psi=\frac{\lambda_2 + \varepsilon\,(\alpha_2\,h + r + \omega_2)}{(1 - h)[\sigma_i (\alpha_2\,h + \omega_2) + \lambda_2 + \varepsilon\,r] + h\,\sigma_s [\sigma_i (\alpha_2\,h + r + \omega_2) + \lambda_2]}.
\]
Using the expression for $h$ in (\ref{h_def}), it follows that the first of these conditions is always satisfied for $\omega_j>0$, and in the limit $\omega_j\to 0$ it turns into $R_0^a>1$. On the other hand, the second condition in the limit $\omega_j\to 0$ turns into $R_0^d< \psi_0$ in agreement with (\ref{R0d_def}).

Finally, the characteristic equation for linearisation near the endemic equilibrium state $E_2^{\omega}=(S_n^*, I_n^*, R_n^*, S_a^*, I_a^*, R_a^*)$ with $\omega_j = \omega$, $\alpha_j= \alpha$ and $\lambda_j= \lambda$, $j = 1, 2, 3$, has the form
\begin{equation}\label{char_eq_end}
\mu(\mu + a_3 + \lambda)(\mu^4 + P_1 \mu^3 + P_2 \mu^2 + P_3 \mu + P_4)=0,
\end{equation}
where
\[
\begin{array}{l}
P_1 = x + y + z + \lambda,\\\\
P_2 = \delta(\phi\delta + \phi a_3 + \lambda) + (x + y)(\lambda + z) +  r(a_9 + a_{10}\,\varepsilon) + xy - (a_3 + a_5)(\lambda + a_4),\\\\
P_3 = \delta(\phi\delta + \phi a_3 + \lambda)(x + y) + (\lambda + z)[xy - (a_3 + a_5)(\lambda + a_4)]\\
\qquad \qquad + ra_9(x + \phi\delta + \lambda) +a_{10}\varepsilon r(y +a_3 + \delta),\\\\
P_4 = \delta (\phi\delta + \phi a_3 + \lambda)\left[xy - (a_3 + a_5)(\lambda + a_4)\right] +  r[a_9x(\lambda + \phi\delta) + a_{10} a_3(\lambda + a_4)] \\
\qquad \qquad +\varepsilon r\left[a_9a_{10}r + \lambda a_9 ( a_3 + a_5) + ya_{10} (\delta + a_3)\right],
\end{array}
\]
and
\[
\begin{array}{l}
\displaystyle{a_9=\frac{(I_n^*+\sigma_i I_a^*)\beta}{N},\hspace{0.5cm}a_{10}=\frac{(I_n^*+\sigma_i\,I_a^*)\,\sigma_s\,\beta}{N}},\\\\
\displaystyle{x = \varepsilon r + a_{10} + \lambda - a_6, \quad y = a_3 + a_7 + r - a_1, \quad z = \phi\delta + \delta + a_3.}
\end{array}
\]
Two of the eigenvalues of the characteristic equation (\ref{char_eq_end}) are $\mu=0$ and $\mu= - (\alpha h + \omega + \lambda)$, so the stability of the endemic steady state $E_2^{\omega}$ is determined by the roots of the quartic
\[
\mu^4 + P_1 \mu^3 + P_2 \mu^2 + P_3 \mu + P_4=0.
\]
Using the Routh-Hurwitz criterion, one can conclude that the steady state $E_2^{\omega}$ is linearly asymptotically stable if and only if the following conditions hold
\begin{equation}
P_4 > 0, \quad P_1 > 0, \quad P_2 > 0 \quad \text{and} \quad P_3\,(P_1\,P_2 -  P_3) > P_1^2\,P_4.
\end{equation}

\begin{figure}[t]
\includegraphics[width = 17cm]{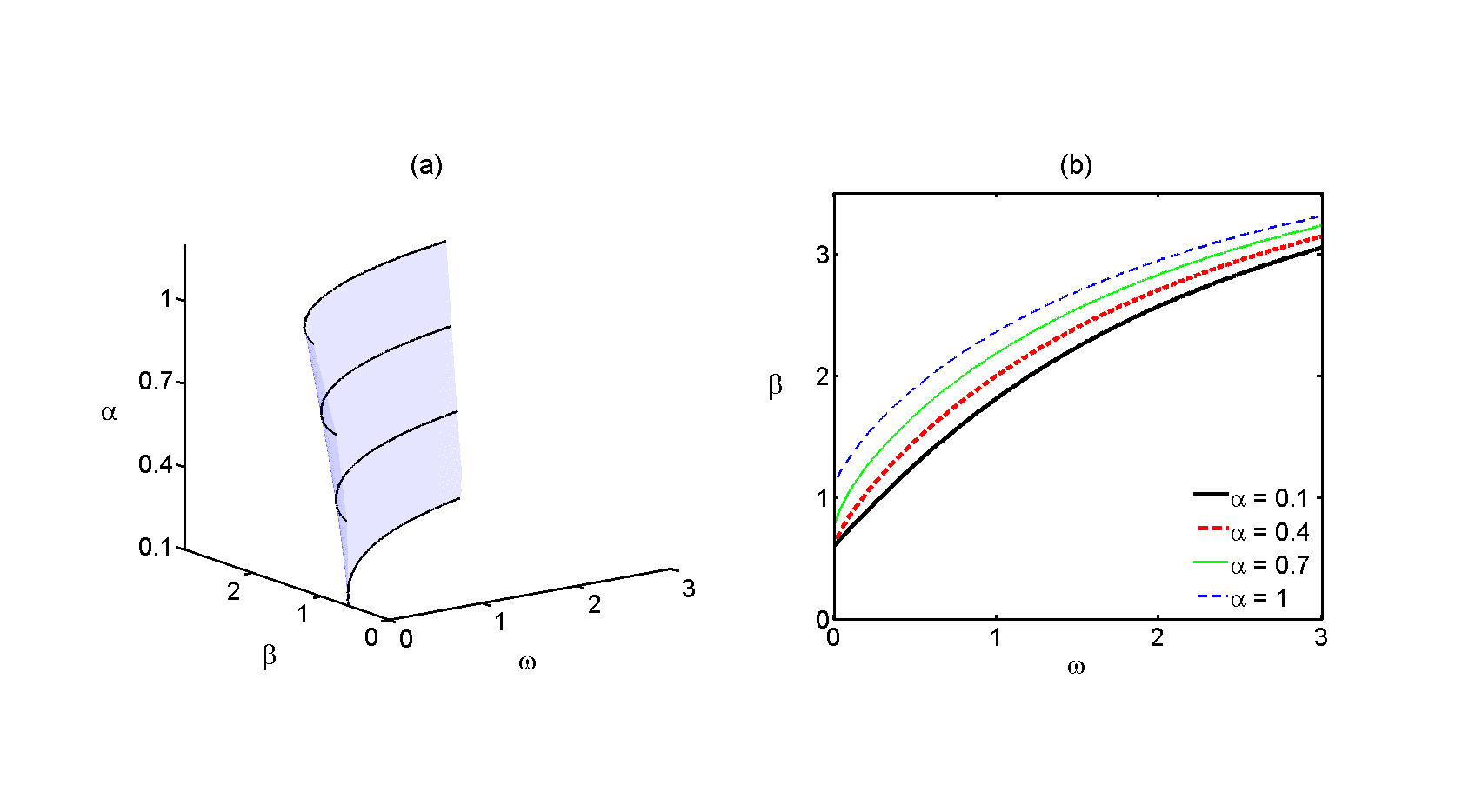}
\vspace{-2cm}
	\caption{Existence and stability of different steady states for $\omega_j = \omega$, $\alpha_j= \alpha$ and $\lambda_j= \lambda$. The disease-free steady state is stable to the right of the surface in (a) and below each curve in (b), and in these parameter regions the endemic steady state is not feasible. To the left of the surface in (a) and above each curve in (b), the disease-free steady state is unstable, while the endemic steady state exists and is stable. Parameter values are $\lambda = 0.6$, $r = 0.6$, $\sigma_i= 0.5$, $\sigma_s=0.5$, $\phi = 0.3$, $\varepsilon=2$, $\delta = 0.4$.}\label{ex_stab1}
\end{figure}

\begin{figure}[t]
	\includegraphics[width = 17cm]{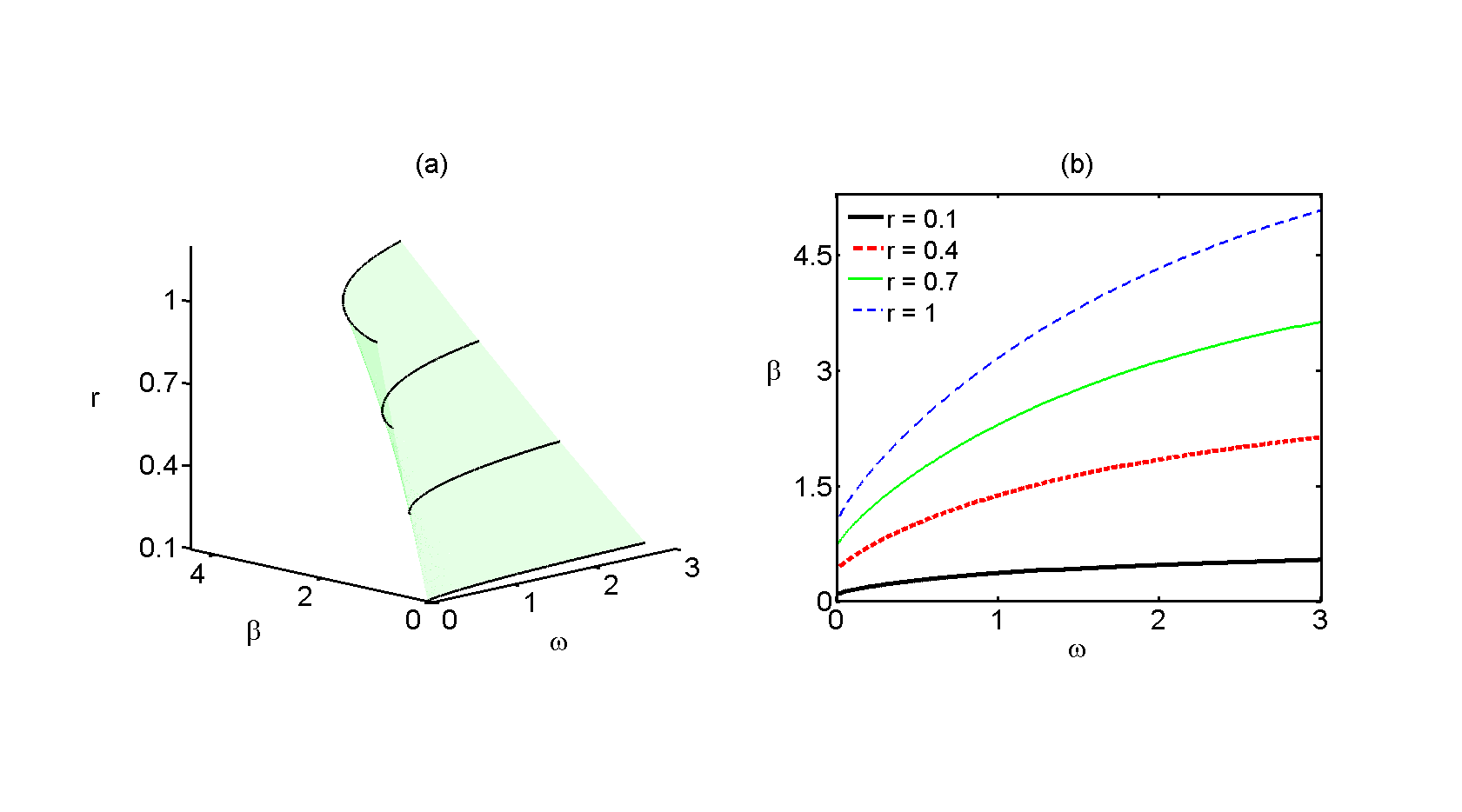}
	\vspace{-2cm}
	\caption{Existence and stability of different steady states for $\omega_j = \omega$, $\alpha_j= \alpha$ and $\lambda_j= \lambda$. The disease-free steady state is stable to the right of the surface in (a) and below each curve in (b), and in these parameter regions the endemic steady state is not feasible. To the left of the surface in (a) and above each curve in (b), the disease-free steady state is unstable, while the endemic steady state exists and is stable. Parameter values are $\alpha=0.4$, $\lambda=0.6$, $\sigma_i=0.5$, $\sigma_s=0.5$, $\phi=0.3$, $\varepsilon=2$, $\delta=0.4$.}\label{ex_stab2}
\end{figure}

Figures~\ref{ex_stab1} and~\ref{ex_stab2} illustrate how the stability of different steady states varies with parameters. Both of these Figures indicate that the endemic steady state is only biologically feasible and stable in the parameter region where the disease-free steady state is unstable. The region of stability of the disease-free steady state increases with $\alpha$ and $\omega$, implying that increasing awareness allows disease eradication and prevents establishment of some steady levels of disease even for higher values of the disease transmission rate $\beta$. Similar effect is observed by increasing the recovery rate $r$, where the disease is eradicated not so much through the spread of awareness, as due to the fact that infected individuals recover faster than they are able to spread the infection. Increasing the rate $\lambda$ of awareness loss naturally has the opposite effect of increasing the parameter region where the endemic steady state is biologically feasible and stable.

\section{Effects of awareness on system dynamics}

In order to get a better understanding of relative effects of different aspects of awareness on determining the stability of different steady states and eventual evolution of the system, we fix three of the four parameters, $\sigma_s$, $\sigma_i$, $\varepsilon$ and $\phi$, to be equal to one, and allow one of them to vary to individually investigate the effect it has on the disease propagation. Qualitative behaviour is similar in all cases considered below in that in the absence of public awareness $(\omega_1=\omega_2=\omega_3=0)$, the epidemic threshold is $R_0^d>1$ for $R_0^a<1$, and $R_0^d>\psi_0$ with $\psi_0=\psi(\omega_j=0)$ for $R_0^a>1$, whereas for $\omega_j>0$, it is given by $R_0^d>\psi$ regardless of the value of $R_0^a$. In the case of $\omega_j=0$ and $R_0^a<1$, the disease is established in the form of a stable disease-endemic steady state $E_1$, while for $R_0^a>1$, and for $\omega_j>0$ and any value of $R_0^a$, the system settles on the stable endemic equilibrium $E_2^\omega$.

In the case of {\it reduced susceptibility}, where $\sigma_i = \varepsilon = \phi = 1$ and $0 \leq \sigma_s < 1$, the epidemic threshold is given by
\begin{equation}\label{psi_rs}
\psi=1 + \frac{h(1 - \sigma_s)}{1-h(1 - \sigma_s)},
\end{equation}
where $h$ was introduced in (\ref{h_def}) and can be equivalently rewritten as
\begin{equation}\label{pub_priv}
h = \frac{1}{2}\,\left(1-\frac{1}{R_0^a} - \frac{\omega_1}{\alpha_1}\right)+\sqrt{\frac{1}{4}\,\left(1-\frac{1}{R_0^a} - \frac{\omega_1}{\alpha_1}\right)^2+\frac{\omega_1}{\alpha_1}}.
\end{equation}
For $\omega_j=\omega=0$, the expression for epidemic threshold reduces to
\begin{equation}\label{R0d_sus}
\psi_0=1 + \frac{(R_0^a-1)(1 - \sigma_s)}{1+(R_0^a-1)\sigma_s}.
\end{equation}
When $\alpha_j=\alpha \to\infty$, this threshold tends to the same limit of $1/\sigma_s$ as the epidemic threshold in a model of Funk et al. \cite{Funk10b}, thus suggesting that when the level of private awareness is much higher than that of public awareness, it is this private awareness that dominates the dynamics, and then it does not really matter whether public awareness extends only to susceptible individuals or to the whole population. However, for intermediate values of $\alpha$, the epidemic threshold in our model depends not only on $R_0^a$ and $\sigma_s$, but also on the ratio of the public $(\omega)$ and private $(\alpha)$ awareness rates as shown in (\ref{pub_priv}), whereas in Funk et al. \cite{Funk10b}, the epidemic threshold was given by (\ref{R0d_sus}) for any value of $\omega$.

\begin{figure}[t]
	\includegraphics[width = 16cm]{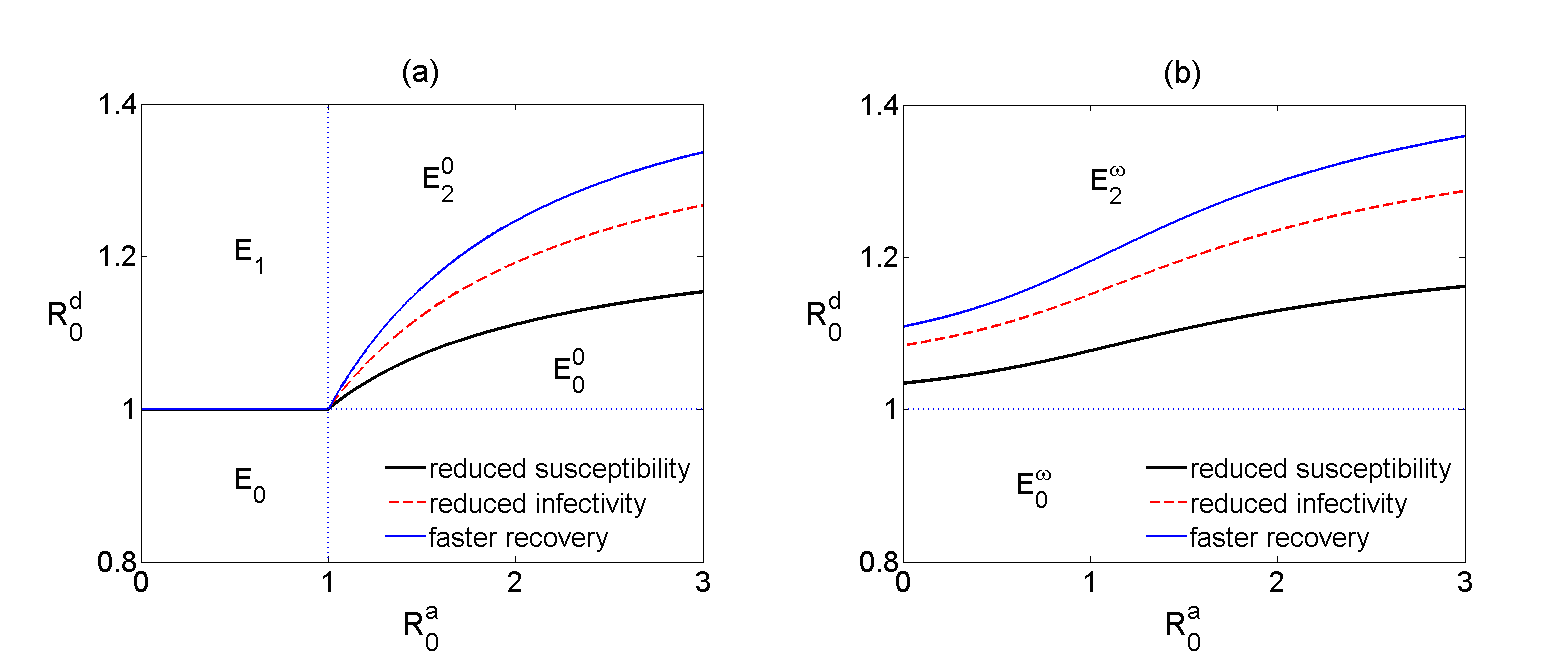}
\caption{Effects of private and public awareness on the spread of infectious diseases for (a) $\omega_1 = \omega_2 = 0$, (b) $\omega_1 = 0.1, \omega_2 = 0.2$. Other parameter values are $\lambda_1 = 0.5, \lambda_2 = 0.4, r = 0.5,  \sigma_i = 0.7, \sigma_s = 0.8, \varepsilon = 1.5, \alpha_1, \alpha_2$ were varied with $\alpha_2 = \alpha_1 + 0.01$. In each case we have indicated a single steady state that is stable in that part of the parameter plane.}\label{fig3}
\end{figure}
\begin{figure}[t]
	\includegraphics[width = 17cm]{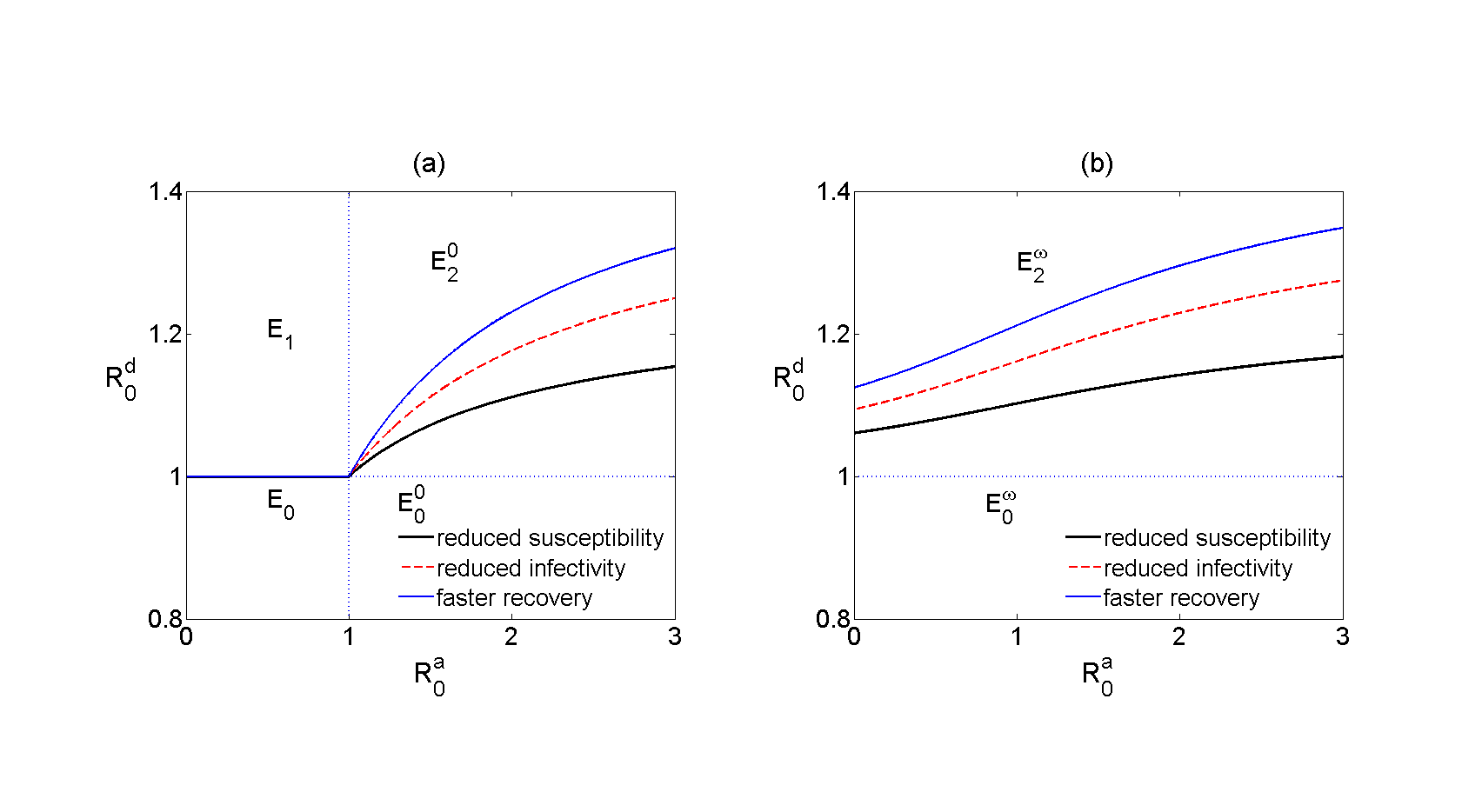}
	\vspace{-1.5cm}
\caption{Effects of private and public awareness on the spread of infectious diseases with $\alpha_j=\alpha$, $\omega_j= \omega$, $\lambda_j= \lambda$, $j = 1,2,3$, for (a) $\omega = 0$, (b) $\omega = 0.2$. Other parameter values are $\lambda = 0.5, r = 0.5,  \sigma_i = 0.7, \sigma_s = 0.8, \varepsilon = 1.5$. In each case we have indicated a single steady state that is stable in that part of the parameter plane.}\label{fig3b}
\end{figure}

When one considers {\it reduced infectivity} where $\sigma_s = \varepsilon = \phi = 1$, and the infective population has its infectivity reduced by a factor $0 \leq \sigma_i < 1$, the epidemic threshold is given by
\[
\psi=1 + \frac{[h(\alpha_2 + r) + \omega_2](1 - \sigma_i)}{\sigma_i[h(\alpha_2 + r) + \omega_2] + \lambda_2 + r(1 - h)},
\]
and similarly to the previous case, it now depends on both types of  awareness and, in fact, it increases with both $\alpha_j$ and $\omega_j \, (j=1,2)$. In the case of {\it faster recovery} with $\sigma_s = \sigma_i = \phi = 1$ and $\varepsilon > 1$, the epidemic threshold becomes
\[
\psi=1 + \frac{[h\,(\alpha_2 + r) + \omega_2]\,(\varepsilon- 1)}{ h\,(\alpha_2 + r) + \omega_2 + \lambda_2 + \varepsilon\,r\,(1 - h)}.
\]
For {\it longer temporary immunity} with $\sigma_s = \sigma_i = \varepsilon = 1$ and $0 \leq \phi < 1$ (average duration of immunity is given by $1/\phi$), the epidemic threshold remains unchanged at $R_0^d>1$. However,  if the awareness of an individual population influences the duration of its immunity $\phi^{-1}$, the fractions of infected and recovered populations in the endemic state can also change \cite{Funk10b}.

Figure~\ref{fig3} illustrates the dependence of epidemic threshold on the values of $R_0^d$ and $R_0^a$ for reduced susceptibility, reduced infectivity and faster recovery. As suggested by the earlier analysis, in the absence of public awareness $(\omega_1=\omega_2=\omega_3=0)$, depending on the values of $R_0^d$ and $R_0^a$ the system can settle on one of the four stable steady states, namely, a disease-free $E_0$, a disease-endemic $E_1$, an awareness-endemic $E_0^0$, or endemic equilibrium $E_2^0$. When the public awareness is present, i.e. $\omega_j>0$, the options are limited to either an awareness-endemic equilibrium $E_0^\omega$, which in this case also plays a role of a disease-free state, and an endemic steady state $E_2^\omega$. Figure~\ref{fig3b} shows the dependence of epidemic threshold on the values of $R_0^d$ and $R_0^a$ for reduced susceptibility, reduced infectivity and faster recovery with $\alpha_j=\alpha$, $\omega_j= \omega$, $\lambda_j= \lambda$ for $j = 1,2,3$. We observe that having unequal values of parameters $\lambda_{1}>\lambda_2$ and $\omega_2>\omega_1$ gives qualitatively similar behaviour of epidemic thresholds to that with equal values of these parameters, though for equal values of parameters, the stability region of the steady state $E_{0}^\omega$ is larger for smaller values of $R_0^a$ and slightly smaller for larger values of $R_0^a$. In the case when $\omega_1=\omega_2$ and $\lambda_1 > \lambda_2$, the results show an increase in the stability region of the steady state $E_0^\omega$ for reduced susceptibility, reduced infectivity and faster recovery. On the contrary, for $\lambda_1 = \lambda_2$ and $\omega_2 > \omega_1$, the stability region for the reduced susceptible has no noticeable change, while for reduced infectivity and faster recovery there is a reduction in the stability region of the steady state $E_0^\omega$.

\begin{figure}[t]
	\includegraphics[width = 17cm]{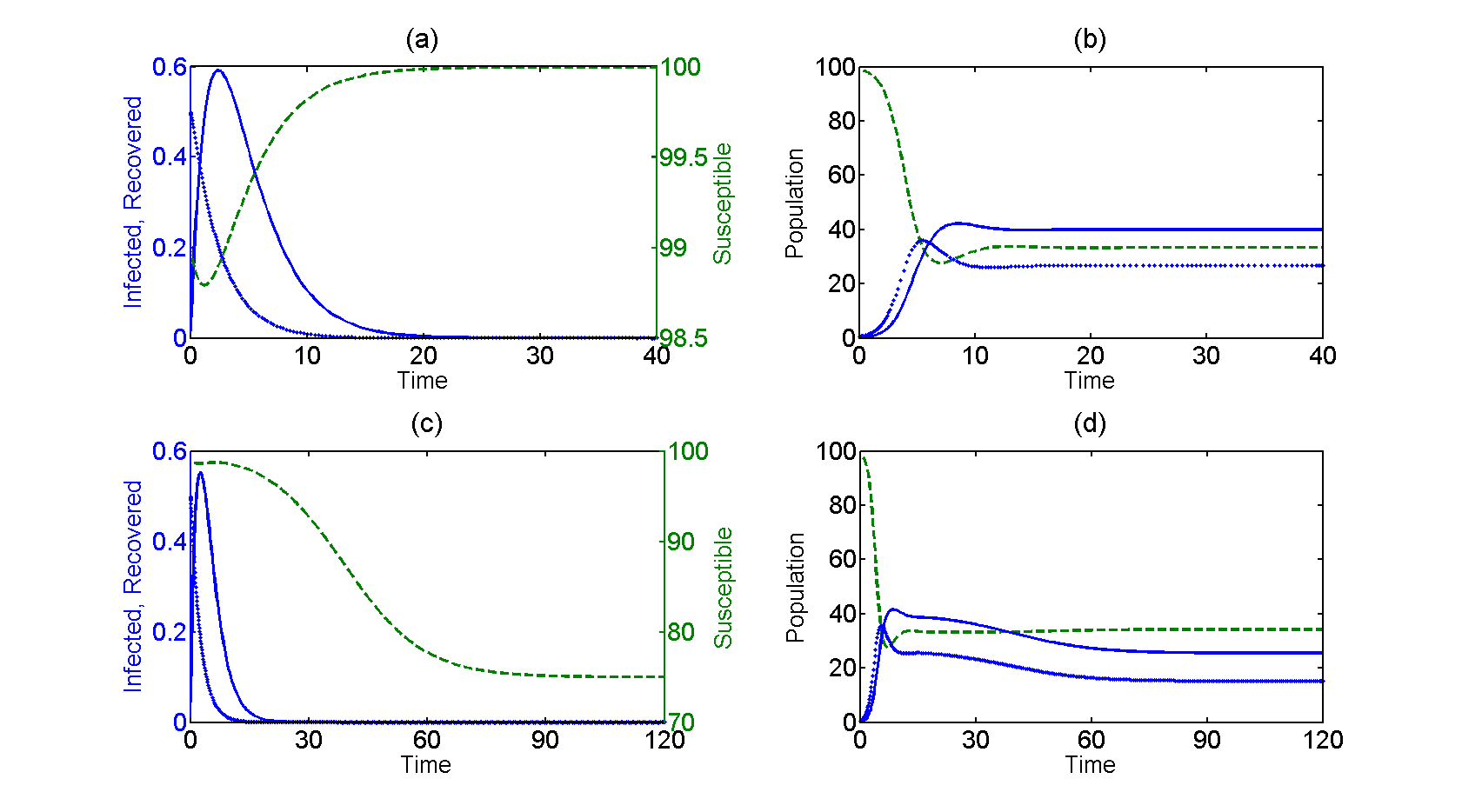}
\vspace{-0.7cm}
	\caption{Steady states in the absence of public awareness $(\omega=0)$. (a) disease-free state $E_0$ with $R_0^d<1, R_0^a <1$ ($r = 1, \beta = 0.6, \lambda = 0.6$) (b) disease-endemic state $E_1$ with $R_0^d > 1, R_0^a <1$ ($r = 0.6, \beta = 1.8, \lambda = 0.6$) (c) awareness-endemic state $E_0^0$ with $R_0^d < \psi_0, R_0^a > 1$ ($r = 1, \beta = 0.6, \lambda = 0.3$) (d) endemic state, $E_2^0$ with $R_0^d >\psi_0, R_0^a > 1$ ($r = 0.6, \beta = 1.8, \lambda = 0.3$). Dashed line denotes $S_n$, dotted line denotes $I_n$, solid line denotes $R_n$. Other parameters are: $\alpha = 0.4, \sigma_i = 0.5, \sigma_s = 0.5, \phi = 0.3,  \varepsilon = 2,  \delta = 0.4,  N = 100$.}\label{NS_om_0}
\end{figure}

In Fig.~\ref{NS_om_0} we show numerical solution of the system (\ref{eqn1}) in the absence of public awareness, i.e. for $\omega_1=\omega_2=\omega_3=0$.
Provided the level of privately acquired awareness is sufficiently small to ensure $R_0^a<1$, and the transmission rate is such that $R_0^d < 1$, after the initial growth, the number of infected individuals decreases, and eventually the system approaches a disease-free steady state $E_0$, as illustrated in Fig.~\ref{NS_om_0}(a). Once the transmission rate exceeds the critical value determined by $R_0^d$, even after the initial outbreak, certain level of disease is maintained in the population, however, all compartments with aware individuals approach zero, thus giving a disease-endemic steady state $E_1$ shown in Fig.~\ref{NS_om_0}(b). \begin{figure}[t]
\hspace{0.5cm}
	\includegraphics[width = 16cm]{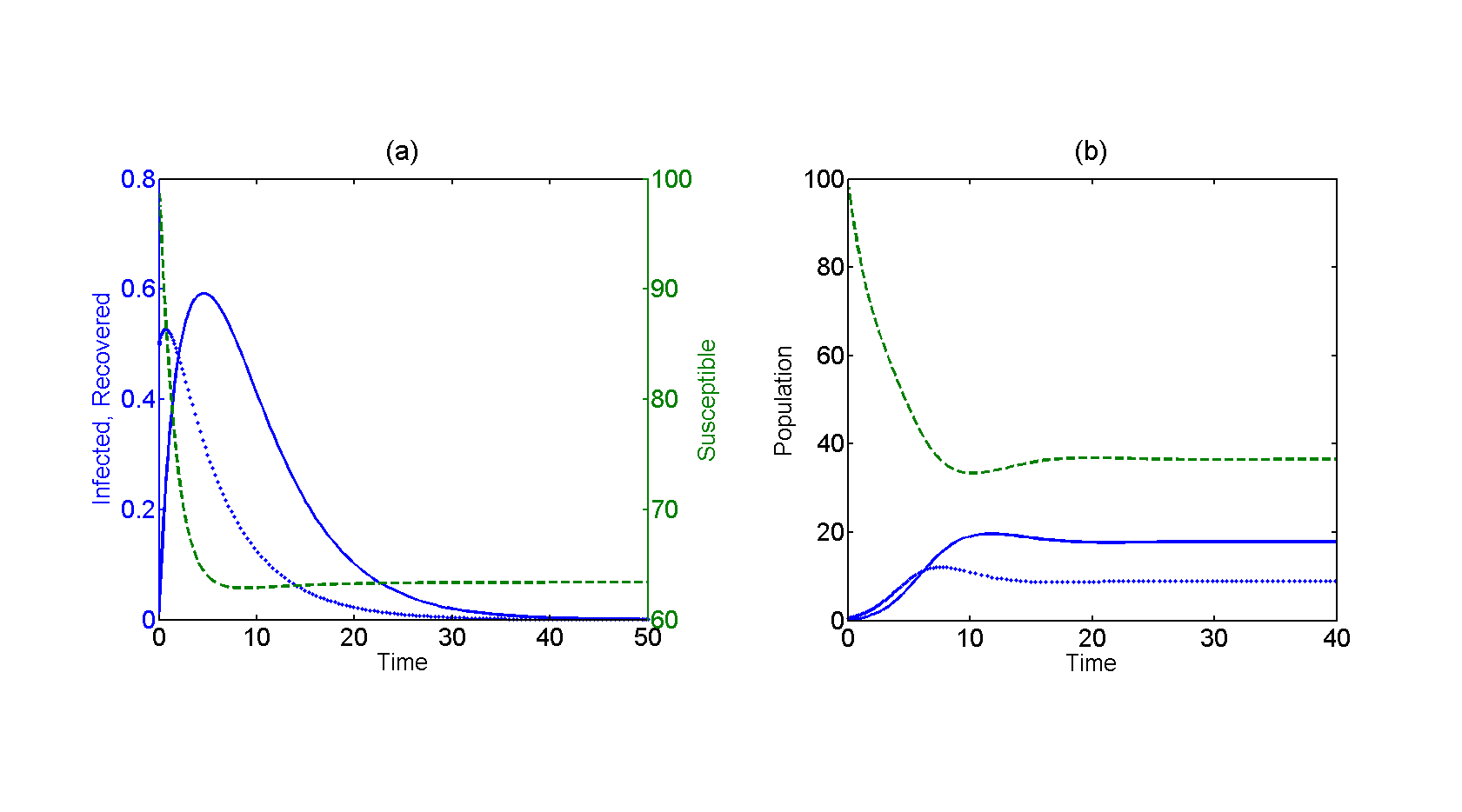}
	\vspace{-1.5cm}
	\caption{Dynamics of infectious disease with public awareness: (a) disease-free state, $\beta = 0.8, R_0^d < \psi$. (b) endemic state, $\beta = 1.8, R_0^d >\psi$. Other parameter values are $\alpha = 0.4, \lambda = 0.6, \omega = 0.2, \sigma_i = 0.5, \sigma_s = 0.5, \phi = 0.3,  \varepsilon = 2,  \delta = 0.4,  N = 100$. Dashed line denotes $S_n$, dotted line denotes $I_n$, solid line denotes $R_n$.}\label{NS_om_pos}
\end{figure}
Figure~\ref{NS_om_0}(c) shows that for sufficiently high private awareness rate, such that $R_0^a>1$, as long as the disease transmission rate $\beta$ is not too high, the population clears the infection, and then the system tends toward an awareness-endemic steady state $E_0^0$. Finally, for higher values of $\beta$, the final state of the system is given by a stable endemic steady state $E_2^0$, as shown in Fig.~\ref{NS_om_0}(d).

In the case of $\omega > 0$ illustrated in Fig.~\ref{NS_om_pos}, there are just two options: the system either approaches a disease-free steady state, whose role is now played by the awareness-endemic steady state $E_0^\omega$ for $R_0^d<\psi$, or it tends to a fully endemic steady state $E_2^\omega$ when $R_0^d>\psi$.

When the values of the parameters $\alpha_j$, $\omega_j$, and $\lambda_j$ are unequal with $\alpha_2 > \alpha_3 > \alpha_1$, $\omega_2 > \omega_3 > \omega_1$, and $\lambda_1 > \lambda_3 > \lambda_2$, numerical simulations suggest a qualitatively similar dynamics of the system, but with a slightly lower peak of unaware infective and recovered individuals, whilst having a slightly higher number of aware susceptible and infected population, and no significant change in the aware recovered population. Similar results are observed if only one or two of the parameters $\alpha$, $\omega$ and $\lambda$ is varied between different groups.

Although we have not rigorously proven global stability of individual steady states, extensive numerical simulations suggest that in each parameter region only one of the steady states of the system is a global attractor, and the system approaches this steady state for arbitrary initial conditions. It is noteworthy that while stability of the disease-free, disease-endemic and awareness-endemic equilibria can change when some parameters are varied, the endemic steady state with all compartments being positive is always stable whenever it is biologically feasible.

\section{Discussion}

This paper has analysed the impact of private and public awareness on the spread of infectious diseases in a human population. The main feature of our model is the possibility of individuals in any of the unaware compartments to become aware both through interactions with aware individuals (regardless of the disease status of the latter), and through a public awareness campaign. This assumption generalises an earlier work of Funk et al. \cite{Funk10b} 
who only accounted for effects of public awareness on infected individuals. Unlike the analysis presented in \cite{Funk10b}, we have been able to obtain analytical expressions for all steady states of the model together with restrictions on parameters that guarantee their biological feasibility, as well as derived analytical conditions for stability of all these equilibria.

Our results show that both private and public awareness have the capacity to reduce the spread of epidemic by increasing the threshold for onset of a stable endemic steady state characterised by persistent infection. Interestingly, unlike some of the earlier models, we have shown that there is an intricate interplay between the two aspects of awareness as illustrated by the dependence of epidemic threshold of $\alpha$ and $\omega$. Quite naturally, the faster people lose awareness (i.e. the larger is the unaware population), the higher is the overall rate of infection as manifested by the disease-endemic state. Conversely, higher recovery rates due to disease awareness leads to a reduction in infected population.
From a more general perspective, the presence of awareness causes corresponding behavioural change in the population, which, in turn, causes the reduction in the size of epidemic outbreaks. Hence, the spread of private awareness or public information campaigns allow one to control or minimise the spread of the disease, whilst they are also helping boost recovery rates for infected individuals. This suggests that information campaigns provide a viable complement if not a replacement for more direct intervention strategies, such as vaccination or quarantine.

In the last few years, some work has been done on modelling the effects of time delays associated with non-instantaneous response of individuals to awareness campaigns \cite{Greenhalgh15,Zhao14,Zuo14,Zuo15}. One direction for extending the work in this paper could be the explicit inclusion of such time delays to more accurately represent the response of individuals to private and public awareness. Another important aspect to be investigated is the development of an optimal vaccination/treatment strategy \cite{Misra15,Wang16} that would utilise behavioural changes associated with the spread of awareness to reduce the amount of vaccine/drug needed to contain an outbreak. For many diseases, an important part of the dynamics is their spatial spread that can also be accompanied with the spreading information about the disease, which can result in very peculiar patterns of disease spread \cite{Bly05,Fer01,Gre01,Wang11}. Whereas in some cases information can improve disease prevention, in others it can result in a public panic thus causing further spread of infection \cite{Ramali}. In this respect, a more advanced version of the model presented in this paper could look into the spatial dynamics of the concurrent spread of disease and infection, as well as optimal strategies for disease containment through targeted spatial dissemination of information and other interventions.

\section*{Acknowledgements} GOA acknowledges the support of the Benue State University through TETFund, Nigeria, and the School of Mathematical and Physical Sciences, University of Sussex.

\end{document}